\documentclass[pre,aps,reprint,english,twocolumn,notitlepage]{revtex4-1}
\usepackage[T1]{fontenc}  \usepackage[latin1]{inputenc}
 \usepackage{babel} \usepackage{txfonts}
 \usepackage{epsfig,epsf,psfrag} 
\usepackage{graphicx} 
\usepackage{subfigure}
\usepackage{pslatex}
\usepackage{epic,eepic} 
\usepackage{color,pstcol}
\usepackage{pstricks}
\usepackage{fancyhdr}  \parindent0em  
 \vspace{-10.0cm}  
\begin{document} \title{Helical thermoelectrics and refrigeration} 
  \author{Arjun Mani}
 \author{Colin Benjamin} \affiliation{School of Physical Sciences, National Institute of Science Education \& Research, HBNI, Jatni-752050, India}
\begin{abstract}
The thermoelectric properties of a three terminal quantum spin Hall (QSH) sample are examined. Inherent helicity of the QSH sample helps to generate a large charge power efficiently. Along  with charge the system can be designed to work as a highly efficient spin heat engine too. The advantage of a helical over a chiral sample is that, while a multiterminal quantum Hall sample can only work  as a quantum heat engine due to broken time reversal(TR) symmetry, a multiterminal QSH system can work effectively both as a charge/spin heat engine as well as a charge/spin refrigerator as TR symmetry is preserved. 
 \end{abstract}
\maketitle
\section{Introduction} Nano-structured materials are attracting a lot of attention due to their large thermopower and low thermal conductances\cite{handbook, benenti}. These large thermo power materials can be used for energy harvesting, i.e., to convert waste heat back into electricity \cite{david}. One further possible use is in refrigeration, i.e., using electrical work to absorb heat from a low temperature region and dumping it in a region at higher temperature\cite{ron}. A two terminal monolayer graphene system has been used as a quantum heat engine(QHE) and refrigerator in presence of strain \cite{arjun}. In two terminal heat engines, the flow of heat energy and electric currents are through the same terminals, so its not possible to control separately the flow of heat and charge current via tuning the transmission function at different terminals. In multi-terminal heat engines, however the separate flow of heat energy and electric current is possible through different terminals. In this work we will discuss a three terminal(3T) quantum spin Hall (QSH) insulator as a QHE and refrigerator. { Quantum spin Hall(QSH) effect is observed at low temperatures in strong spin-orbit coupling systems like HgTe/CdTe quantum well structure.  Similar to the quantum Hall (QH) effect, here too, 1D gap less edge states appear. These edge states are spin-momentum locked, i.e., if spin up electron is moving in one direction then spin down electron is moving in the opposite direction at one edge of the sample, and at the other edge vice-versa. These are called helical edge states. 2D edge/surface states are also included in QSH effect but materials are Bi$_2$Se$_3$, Bi$_2$Te$_3$, Bi$_{1-x}$Sb$_x$ etc. Since our work deals with 1D QSH edge modes, our candidate materials are HgTe/CdTe quantum well structures. Using the helical properties of the 1D edge modes we have designed a powerful quantum heat engine as well as a quantum refrigerator.} In Refs.~\cite{sothmann, sothmann2} a 3T quantum Hall (QH) system is shown to work as a QHE with the aid of quantum interference or quantum point contacts (QPC). These multi-terminal QH heat engines have broken TR symmetry and thus have either the Seebeck coefficient finite and Peltier coefficient zero or vice-versa due to the presence of chiral edge modes. The asymmetric parameter(AP)- ratio of Seebeck to Peltier coefficient, in these models is therefore either zero or infinity. AP is intimately related to the working of a heat engine as refrigerator. The fact that AP is either zero or infinity reduces the ability of QH heat engines to be used as a refrigerator, see Ref.~\cite{brandner}. In contrast for a QSH system TR symmetry is not broken and thus AP is unity, which implies that the upper bound of coefficient of performance (COP) of a QSH refrigerator is equal to the Carnot efficiency of the refrigerator. We will discuss our model of a QSH 3T system, shown in Fig.~1, both as QHE as well as refrigerator working at full power. 
{ Due to the quantum effects and in the non-linear transport regime there is an upper limit to how much heat energy can be carried by each channel/edge mode, see Ref. \cite{robert}. As a result, it also limits the efficiency achieved at maximum power by any heat engine irrespective of whether it is two/three terminal heat engine or TR symmetry is broken or not. Though this kind of bound will not affect our results as we are in linear transport regime where the temperature difference applied between the two terminals is small and the heat energy carried by each edge mode will always be less than this upper bound.}

The manuscript is organized as follows. We next discuss the theory required to explain the QSH heat engine and refrigerator, working in both charge as well as spin domains. Next, we discuss our model, which consists of a 3T QSH system with energy dependent transmissions through two constrictions X,Y (Fig.~1) which can be designed  by QPC's or antidots \cite{sothmann2}. Following this we discuss the results of our paper with few plots of the thermopower, charge/spin power and efficiencies for both QSH heat engine and refrigerator. Finally, we discuss the experimental realization of our work along with Tables I, II, comparing our results with similar proposals for heat engines and refrigerators.
\begin{figure}
\includegraphics[width=0.5\textwidth]{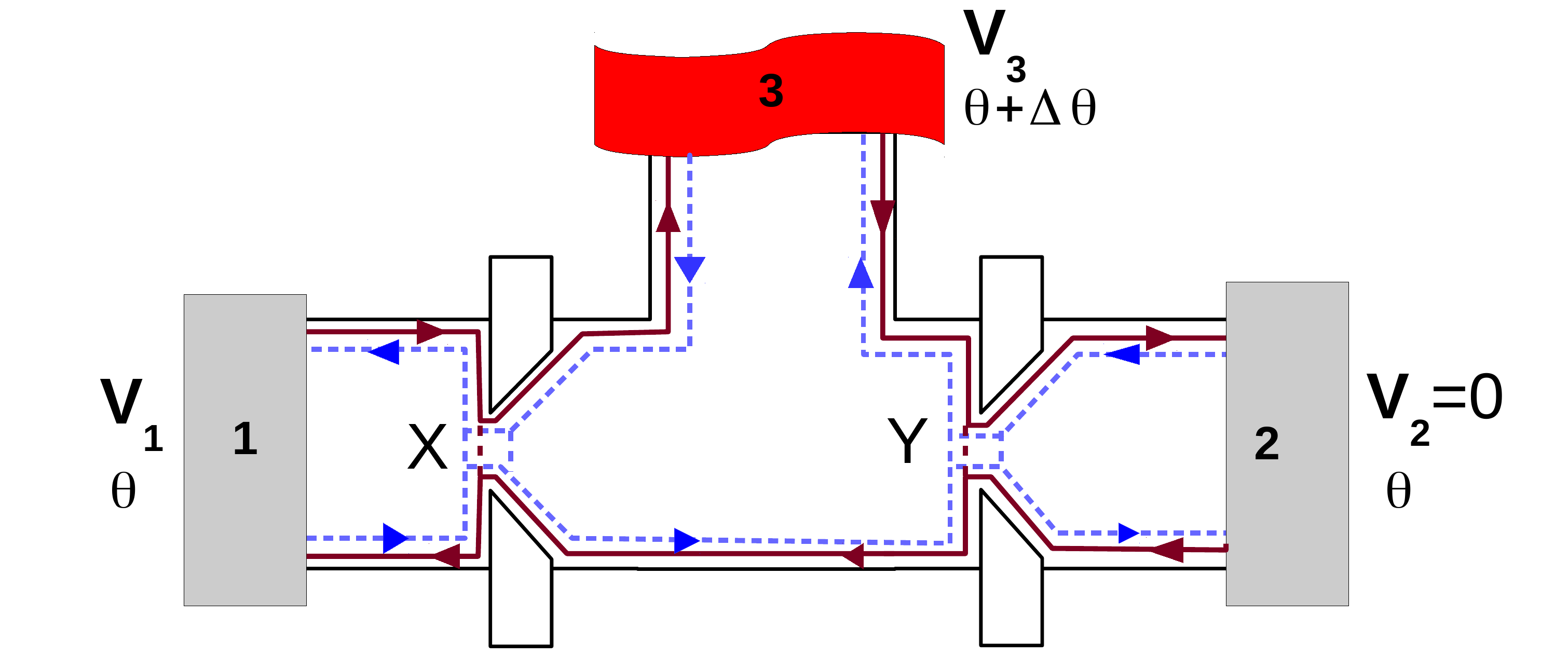}
\caption{3T QSH thermoelectric system. Blue dashed line represents spin up and maroon solid line represents spin down edge mode. Voltage bias $\Delta V$ is applied between terminals 1 and 2. Thermal gradient is applied at terminal 3 which acts as a voltage probe too.}
\end{figure}

{\section{Theory} }
\subsection{ QSH heat engine}
We dwell here on a 3T QSH thermoelectric system. For simplicity, we have considered only one spin up edge mode shown by blue dashed line and one spin down edge mode by maroon solid line, see Fig.~1. The terminals 1 and 2 are at same temperature $\theta$, while the terminal $3$ is at a higher temperature $\theta_3=\theta+\Delta \theta$ with respect to the other terminals. We describe the problem via Landauer-Buttiker formalism, i.e., the electric and heat currents transported from one terminal to another are defined via the transmission probabilities as long as we are in the linear response regime\cite{ramshetti}. In linear response regime, the electric ($I^{e,s}_i$) and heat currents ($I^{h,s}_i$) can be written in terms of the driving forces, i.e., bias voltage and temperature difference, as \cite{sothmann2, ramshetti}-
{\small
\begin{equation}
  \left( {\begin{array}{c}
   I^{e,s}_i  \\
   I^{h,s}_i  \\
  \end{array} } \right)= \frac{1}{h} \sum_j \int_{-\infty}^{\infty} dE [\delta_{ij}-T^s_{ij}(E)] (-\frac{df}{dE}) \left( {\begin{array}{cc}
 e^2 & eE/\theta  \\
   eE & E^2/\theta \\
  \end{array} } \right)  \left( {\begin{array}{c}
  \Delta V_j  \\
   \Delta \theta_j  \\
  \end{array} } \right)
\end{equation}
}
where $T^s_{ij}(E)$ is the energy dependent transmission from terminal $j$ to $i$ for spin $s=\uparrow/\downarrow$ electrons, $f$ is the Fermi-Dirac distribution with Fermi energy $E_F=0$, and $I^{e,s}_i, I^{h,s}_i$ define the electric and heat currents at terminal $i$ for spin $`s$' electrons and $\Delta \theta_3=\Delta \theta$ with $\Delta \theta_1=\Delta \theta_2=0$ is the thermal bias applied only at terminal $3$. Since terminal $3$ is a voltage probe, electric charge current, $I^e_{ch,3}=I^{e,\uparrow}_3+I^{e,\downarrow}_3$, through it is zero. From conservation of current we thus have $I^e_{1,ch}=-I^e_{2,ch}$, and as temperature difference is applied only at terminal $3$, we can rewrite Eq.~(1) in terms of Onsager coefficients, see \cite{ramshetti} as-
{ \begin{equation}
  \left( {\begin{array}{c}
   I^{e,s}  \\
   I^{h,s}_3  \\
  \end{array} } \right)= \left( {\begin{array}{cc}
 L^s_{eV} & L^s_{e\theta}  \\
   L^s_{hV} & L^s_{h\theta} \\
  \end{array} } \right)  \left( {\begin{array}{c}
  \Delta V  \\
   \Delta \theta_3  \\
  \end{array} } \right),
\end{equation}   }
where $L^s_{eV}=G^s$ is the electric and $L^s_{h\theta}$ is the thermal conductance respectively for spin $`s$' electrons, while the off-diagonal elements are the thermoelectric responses. Since in our work, we do not have any spin-flip scattering, from Eq.~(2), one can define charge/spin Seebeck ($S_{ch/sp}$) and Peltier coefficients ($P_{ch/sp}$) as-
{ \begin{eqnarray}
S_{ch}=\frac{S^\uparrow+S^\downarrow}{2}, \quad S_{sp}=S^\uparrow-S^\downarrow, \text{ with } S^s=\frac{L^s_{e\theta}}{L^s_{eV}}, \\
P_{ch}=\frac{P^\uparrow+P^\downarrow}{2}, \quad P_{sp}=P^\uparrow-P^\downarrow,\text{ with }P^s=\frac{L^s_{hV}}{L^s_{eV}}.
\end{eqnarray}}
Summing over spin $s$ of the electrons, in Eq.~2, we can write the charge $I^e_{ch}=I^{e,\uparrow}_1+I^{e,\downarrow}_1$, spin $I^e_{sp}=I^{e,\uparrow}_1-I^{e,\downarrow}_1$ electric currents at terminal $1$ and the heat $I^h_{ch}=I^{h,\uparrow}_3+I^{h,\downarrow}_3$ current at terminal $3$ in terms of the driving forces $ V_{ch}$, $ V_{sp}$ and $\Delta \theta$ as follows\cite{Bauer}-  
\begin{eqnarray}
  \left( {\begin{array}{c}
   I^{e}_{ch}  \\
   I^{e}_{sp}  \\
    I^{h}_{ch}\\
  \end{array} } \right)&=& \left( {\begin{array}{ccc}
 G_{ch}&G_{sp}& L^{+}_{e\theta}  \\
  G_{sp} & G_{ch}&L^{-}_{e\theta} \\
   L^{+}_{hV}&L^{-}_{hV}&L^{+}_{h\theta}\\
  \end{array} } \right)  \left( {\begin{array}{c}
   V_{ch}  \\
  V_{sp}/2  \\
  \Delta \theta\\
  \end{array} } \right),
  \end{eqnarray}   
where $V_{ch}=\frac{V^{\uparrow}+V^{\downarrow}}{2}$ and $V_{sp}=V^{\uparrow}-V^{\downarrow}$ are the charge and spin voltages at terminal $1$, $G_{ch}=G^{\uparrow}+G^{\downarrow}$ and $G_{sp}=|G^{\uparrow}-G^{\downarrow}|$ are the charge and spin conductances respectively. The thermoelectric responses are defined as $L^\pm_{k}=L^{\uparrow}_{k}\pm L^{\downarrow}_{k}$, where $k=hV$, $h\theta$ or $e\theta$. In our setup we apply only a charge voltage bias $V_1-V_2=\Delta V=V_1$, thus $V_{sp}=0$ and $V_{ch}=V_1$. This gives the output power for charge current at terminal $1$ as- 
\begin{equation}
\mathcal{P}_{ch}=I^e_{ch} V_1= G_{ch}V_1^2 +L^{+}_{e\theta}V _1\Delta \theta.
\end{equation}
 The maximum charge output power can be calculated by differentiating ${\mathcal{P}}_{ch}$ with respect to $V_1$ and equating it to zero, $\frac{d\mathcal{P}_{ch}}{dV_1}=0$. This gives the maximum output charge power at $V_1=-\frac{L^{+}_{e\theta}}{2G_{ch}}\Delta \theta$. Similarly, the output power for spin current- 
 \begin{equation}
 \mathcal{P}_{sp}=I^e_{sp}V_1=G_{sp}V^2_1+L^-_{e\theta}V_1\Delta \theta,
 \end{equation}
  can also be set to maximum via $\frac{d\mathcal{P}_{sp}}{dV_1}=0$, which gives the maximum at $V_1=-\frac{L^{-}_{e\theta}}{2G_{sp}}\Delta \theta$.
The maximum charge/spin output power at terminal $1$ can thus be calculated from Eqs.~(6) and (7) as-
\begin{eqnarray}
{\mathcal{P}}^{max}_{ch}=\frac{1}{4}\frac{(L^{+}_{e\theta})^2}{G_{ch}}(\Delta \theta)^2\text{ and } {\mathcal{P}}^{max}_{sp}=\frac{1}{4}\frac{(L^{-}_{e\theta})^2}{|G_{sp}|}(\Delta \theta)^2.
\end{eqnarray}
Following from Eq.~(8), the charge/spin efficiency at that maximum charge/spin power can be calculated by substituting $V_1=-\frac{L^{+}_{e\theta}}{2G_{ch}}\Delta \theta$ for charge currents and $V_1=-\frac{L^{-}_{e\theta}}{2G_{sp}}\Delta \theta$ for spin currents in expressions for ${\mathcal{P}}^{max}_{ch}$ and $P^{max}_{sp}$ as follows-
\begin{eqnarray}
\eta({\mathcal{P}}^{max}_{ch})&=&\frac{\mathcal{P}^{max}_{ch}}{I^h_{ch}}=\theta\frac{\eta_c}{2}\frac{(L^{+}_{e\theta})^2}{2G_{ch}L^+_{h\theta}-L^+_{e\theta}L^+_{hV}},\nonumber\\ \eta({\mathcal{P}}^{max}_{sp})&=&\frac{\mathcal{P}^{max}_{sp}}{I^h_{ch}}=\theta\frac{\eta_c}{2}\frac{(L^-_{e\theta})^2}{2G_{sp}L^+_{h\theta}-L^-_{e\theta}L^+_{hV}}
\end{eqnarray}
Eqs.~(8, 9) are the main working formulas for the QSH heat engine. Next we explore how to turn our model into a quantum refrigerator for both charge as well as spin.

 {\subsection{ QSH refrigerator}} For our model depicted in Fig.~1 to work as a quantum refrigerator, first we need to define the co-efficient of performance (COP)\cite{brandner}. COP is the ratio of heat current extracted by the system from cooler terminal to the electrical work done on the system. Here, the terminals $1$ and $2$ are both at the same temperature, i.e., cooler than terminal $3$. So, heat is absorbed from terminals $1$ and $2$ and dumped into terminal $3$. Mathematically, COP is defined as- $\eta^r_{ch}=\frac{J^Q}{W_{ch}}$ for charge currents, wherein $J^Q=I^h_{ch}=-\sum_s[(I^{h,s}_1+I^{h,s}_2)]=\sum_sI^{h,s}_3$, see Eq.~(1). The charge output power $W_{ch}=P_{ch}$ is the electrical work done on the system via charge currents with $I^e_{ch}$ defined as in Eq.~(5). Similarly in case of spin, we can define COP(spin) given by\cite{arjun}- $\eta^r_{sp}=\frac{J^Q}{W_{sp}}$, where $W_{sp}=I^e_{sp}V_1$ is the spin work done on the system via spin currents with $I^e_{sp}$ is given in Eq.~(5). 
COP of the system can be set to maximum for given charge/spin currents by allowing for $\frac{d\eta^r_{ch(sp)}}{dV}=0$, which is maximum for charge current (considering $J^Q<0$ and $W_{ch}>0$) at \cite{benetti}- 
\begin{equation}
 V=-\frac{L^+_{h\theta}}{L^+_{hV}}\left(1+\sqrt{\frac{det \mathbb{L}^+}{G_{ch}L^+_{h\theta}}}\right)\Delta \theta, 
\mbox {where } \mathbb{L}^+=\left({\begin{array}{cc}G_{ch}&L^+_{e\theta}\\L^{+}_{hV}&L^+_{h\theta}\end{array}}\right) \end{equation} 
and $det\mathbb{L}^+$ refers to determinant of matrix $\mathbb{L}^+$. The maximum COP and the cooling power $J^Q$ for the charge currents are -
\begin{eqnarray}
\eta^{r,max}_{ch}=\frac{\eta^r_c}{x}\frac{\sqrt{y+1}-1}{\sqrt{y+1}+1},&& \quad \text{ with } y=\frac{L^+_{hV}L^+_{e\theta}}{det \mathbb{L}^+},\quad \text {} x=\theta\frac{ L^+_{e\theta}}{ L^+_{hV}},\nonumber\\
\quad \text {and  } J^Q(\eta^{r,max}_{ch})&=&L^+_{h\theta}\left(\sqrt{\frac{det \mathbb{L}^+}{G_{ch}L^+_{h\theta}}}\right)\Delta \theta,
\end{eqnarray}
while COP for spin currents is maximum at -
\begin{equation}
 V=\frac{L^+_{h\theta}}{L^+_{hV}}\left(-1-\sqrt{\frac{det \mathbb{L}^-}{G_{sp}L^+_{h\theta}}}\right)\Delta \theta,
\mbox {where } \mathbb{L}^-=\left({\begin{array}{cc}G_{sp}&L^-_{e\theta}\\L^{+}_{hV}&L^+_{h\theta}\end{array}}\right)
\end{equation} 
and $det\mathbb{L}^-$ refers to determinant of matrix $\mathbb{L}^-$. The maximum COP and cooling power at that maximum COP for spin current is-
\begin{eqnarray}
\eta^{r,max}_{sp}=\frac{\eta^r_c}{x}\frac{\sqrt{y'+1}-1}{\sqrt{y'+1}+1},&& \quad \text{ with}\quad y'=|\frac{L^+_{hV}L^-_{e\theta}}{det \mathbb{L}^-}|,\nonumber\\
\quad \text {and  } J^Q(\eta^{r,max}_{sp})&=&L^+_{h\theta}\left(\sqrt{\frac{det \mathbb{L}^-}{G_{sp}L^+_{h\theta}}}\right)\Delta \theta,
\end{eqnarray}
where, $\eta^r_c=\theta/\Delta \theta$ is the Carnot efficiency of refrigerators. Our model can work both as a quantum heat engine as well as a quantum refrigerator as it does not break TR symmetry. This is a major advantage of our work in comparison to quantum Hall heat engines which are difficult to convert for refrigeration. For systems with broken TR symmetry the asymmetry parameter(AP) $x=\theta\frac{ L^+_{e\theta}}{L^+_{hV}}$(ratio of Seebeck to Peltier coefficient), deviates from unity. The more AP deviates from unity, more the upper bound of COP goes below the Carnot efficiency $\eta^r_c$\cite{brandner}.

\section{Model} A 3T QSH bar is shown in Fig.~1. The transmissions between the terminals, is modulated by constrictions at X, Y. The transmission through these  constrictions is energy dependent, which is the main criteria to get a finite thermoelectric response. Here, we discuss two kinds of transmission (see \cite{sothmann2})-a) \underline{ QPC like:} the transmission below a certain energy is zero, and above a particular energy is unity and in between it is partially transmitting, mathematically, $\mathcal{T}^{QPC}_l(E)=[1+exp(-2\pi(E-E_l)/\hbar \omega_{0})]^{-1}$  and b) \underline{ resonant tunneling like:} only at a particular energy range the transmission is finite, otherwise zero, mathematically, $\mathcal{T}^{RES}_l(E)=\Gamma_l^2[\Gamma_l^2+4(E-E_l)^2]^{-1}$. Here, $E_l$ is the position of the step at constriction $l=X,Y$, while $\omega_{0}$ and $\Gamma_l$ are the width of the same for QPC and resonant tunneling respectively. The first kind of transmission is present in case of QPC constrictions, and the second kind is present in case of antidot constrictions \cite{sothmann2}. Depending on what kind of transmission is present at which constriction, there are four possible configurations. Configuration 1 consists of two QPC's at X and Y, configuration 2 consists of a QPC at  X and an antidot (resonant tunneling) at Y. Configuration 3 consists of an antidot at X and a QPC at Y while configuration 4 consist of two antidots at X, Y. 
\begin{figure}
 \centering    \subfigure[]{\includegraphics[width=0.23\textwidth]{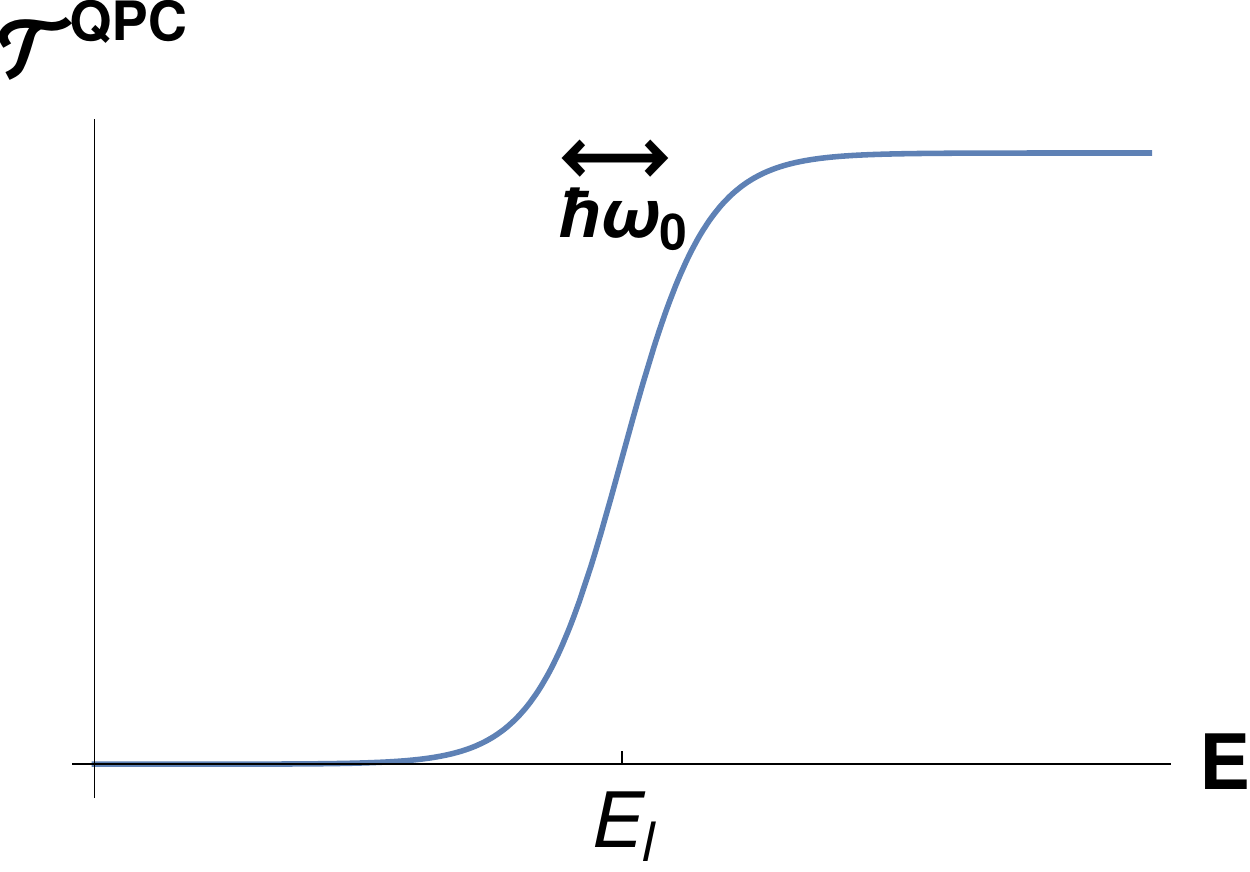}}
 \centering    \subfigure[]{\includegraphics[width=0.23\textwidth]{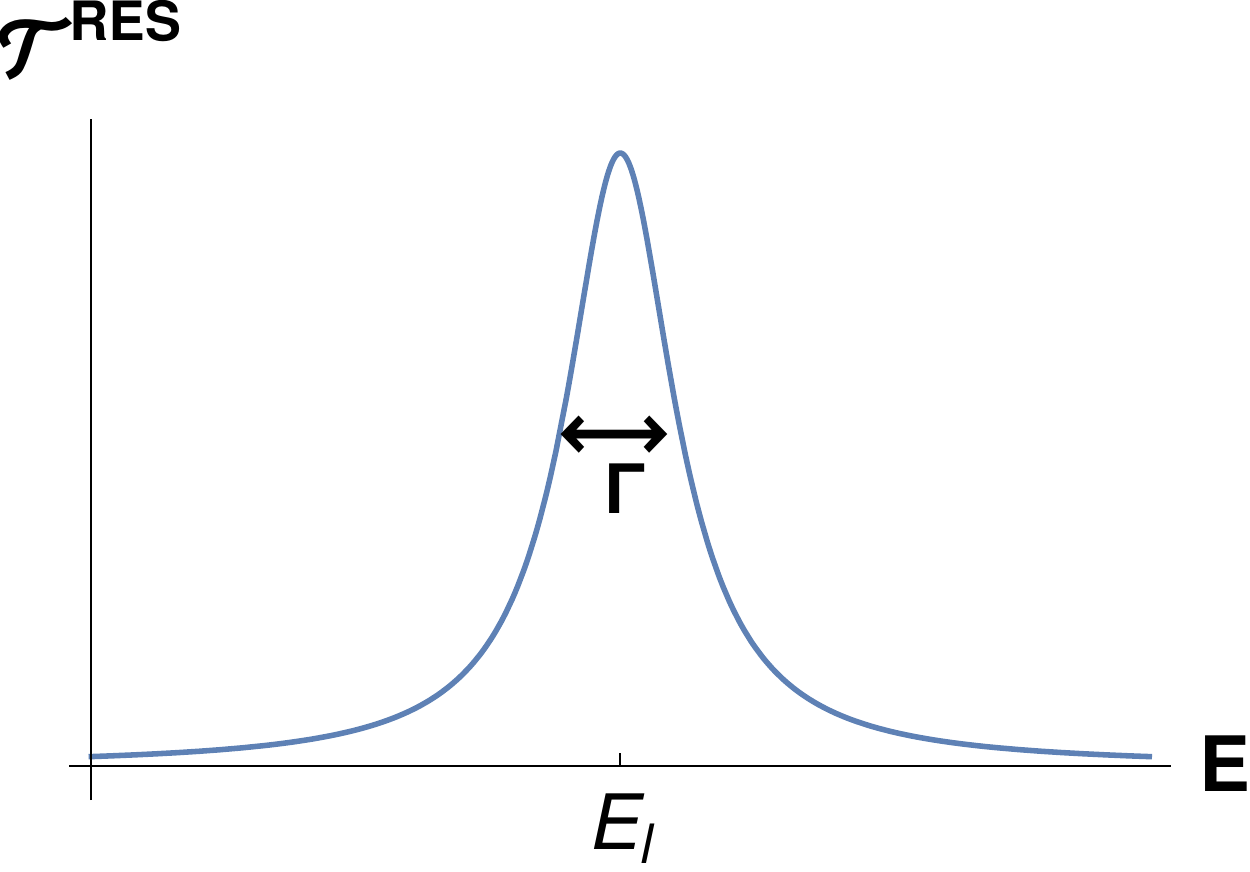}}
 \vskip -0.2 in
\caption{Two types of energy dependent transmission-a) QPC type- described by a saddle point potential, b) resonant tunneling type- due to the presence of an antidot.}
\end{figure}
To calculate maximum power and efficiency at that maximum power, first we need to calculate the conduction $G_s$ and Seebeck coefficient $S_s$ for spin $s$ electrons. The thermoelectric response is generated due to the energy dependent transmission through the QPC's/antidots between the terminals\cite{sothmann3} and is calculated below. The conduction of spin up and spin down electrons can be calculated in a 3T QSH bar following Landauer-Buttiker formalism. For a multi-terminal setup with thermoelectric transport, the spin dependent electric and heat currents are given below\cite{sanchez}-
\begin{eqnarray}
I^{e,s}_i=\sum_jG^s_{ij}V_j+\sum_j L^s_{ij,e\theta}\Delta \theta_j, \nonumber\\
I^{h,s}_i=\sum_jL^s_{ij,hV}V_j+\sum_j L^s_{ij,h\theta}\Delta \theta_j,
\end{eqnarray} 
where, $G^s_{ij}=\frac{e^2}{h}\int_{-\infty}^{\infty} dE [M^s_i\delta_{ij}-T^s_{ij}(E)](-\frac{df}{dE})$, $L^s_{ij,e\theta}=\frac{e}{h\theta}\int_{-\infty}^{\infty} dE (E-\mu) [M^s_i\delta_{ij}-T^s_{ij}(E)](-\frac{df}{dE})=L^s_{ij,hV}/\theta$ and $L^s_{ij,h\theta}=\frac{1}{\theta h}\int_{-\infty}^{\infty} dE (E-\mu)^2 [M^s_i\delta_{ij}-T^s_{ij}(E)](-\frac{df}{dE})$ with $M^s_i=$the no. of edge modes at contact `$i$' for spin `$s$' electron.
The derivation of spin conductances $G^s_{ij}$ and spin thermoelectric coefficients $L^s_{ij,e\theta}$ and their relation to the constriction conductances $ G_l$ and thermopower $S_l$ with $l=X, Y$ are explained in Appendix A. Thus, Eq.~(14) for electric currents reduces to:
\begin{eqnarray}
 I^{e,\uparrow}_1=G^\uparrow V_1+L^{\uparrow}_{e\theta}\Delta \theta,\quad\text{and}\quad
  I^{e,\downarrow}_1=G^\downarrow V_1+L^{\downarrow}_{e\theta}\Delta \theta.
\end{eqnarray}
with,
\begin{eqnarray}
G^\uparrow&=&\frac{G_X(2G_Y-J_1)}{2(G_X+G_Y-J_1)}, \qquad G^\downarrow=\frac{2G_XG_Y+G_XJ_1-J_1^2}{2(G_X+G_Y-J_1)},\nonumber\\
L^{\uparrow}_{e\theta}&=&\frac{G_XG_Y}{G_X+G_Y-J_1}(S_Y-S_X)+\frac{G_X}{G_X+G_Y-J_1}(J_1S_X-J_2),\nonumber\\
L^{\downarrow}_{e\theta}&=&\frac{G_XG_Y}{G_X+G_Y-J_1}(S_Y-S_X)-\frac{G_Y}{G_X+G_Y-J_1}(J_1S_Y-J_2),\nonumber\\
\end{eqnarray}
for spin up and down electric currents. Further $G_l=\frac{e^2}{h}\int_{-\infty}^{\infty}dE\mathcal{T}_l(E)(-\frac{df}{dE})$, $J_n=A_n\int_{-\infty}^{\infty}dE(E-\mu)^{n-1}  \mathcal{T}_X(E) \mathcal{T}_Y(E) (-\frac{df}{dE})$ with $A_1=\frac{e^2}{h}$, $A_2=\frac{e}{\theta h}$, $A_3=\frac{1}{\theta h}$ and $S_l=\frac{e}{\theta hG_l}\int_{-\infty}^{\infty}  dE (E-\mu)(-\frac{df}{dE})\mathcal{T}_l(E)$ is the thermopower across the QPC's/antidots at constriction `$l$'.  The first term in the thermoelectric responses ($L^s_{e\theta}, s=\uparrow,\downarrow$) Eq.~(16) is proportional to the difference between the thermopower generated at the two constrictions. The second term is related to the coherent transport between the respective terminals and the sign of this term is related to the helicity of the different spins. Spin up electrons are moving in counter clockwise direction, which is opposite to that of spin down electrons which are moving in clockwise direction. So, different spins have opposite effect on the thermoelectric responses as shown in the second term. Similar to the electric currents, for spin up and down heat currents we get-
\begin{eqnarray}
I^{h,\uparrow}_3&=&L^\uparrow_{hV}V_1+L^\uparrow_{h\theta}\Delta \theta,\quad
I^{h,\downarrow}_3=L^\downarrow_{hV}V_1+L^\downarrow_{h\theta}\Delta \theta, \\
&& \mbox{where, } L^\uparrow_{hV}=\theta L^\downarrow_{e\theta}, \qquad L^\downarrow_{hV}=\theta L^\uparrow_{e\theta}, \text{ and}\nonumber\\
L^{\uparrow}_{h\theta}&=&L^{\downarrow}_{h\theta}=(N_1+N_2-J_3)-\theta\frac{(G_XS_X+G_YS_Y-J_2)^2}{(G_X+G_Y-J_1)}.
\end{eqnarray}
The derivation of thermoelectric responses and their relation to the conductances and thermopower across constrictions X, Y are shown in Appendix B. $N_l=\frac{1}{\theta h}\int_{-\infty}^{\infty}  dE \mathcal{T}_l(E)(E-\mu)^2(-\frac{df}{dE})$ is the thermal conductance across the QPC/antidot at constriction `$l$'. From Eq.~(18) we see that $L^+_{hV}=\theta L^{+}_{e\theta}$, which implies that TR symmetry is preserved in 3T QSH systems unlike in 3T QH systems, see Ref.~\cite{sothmann2}.
 \begin{figure}
\centering {\includegraphics[width=0.4\textwidth]{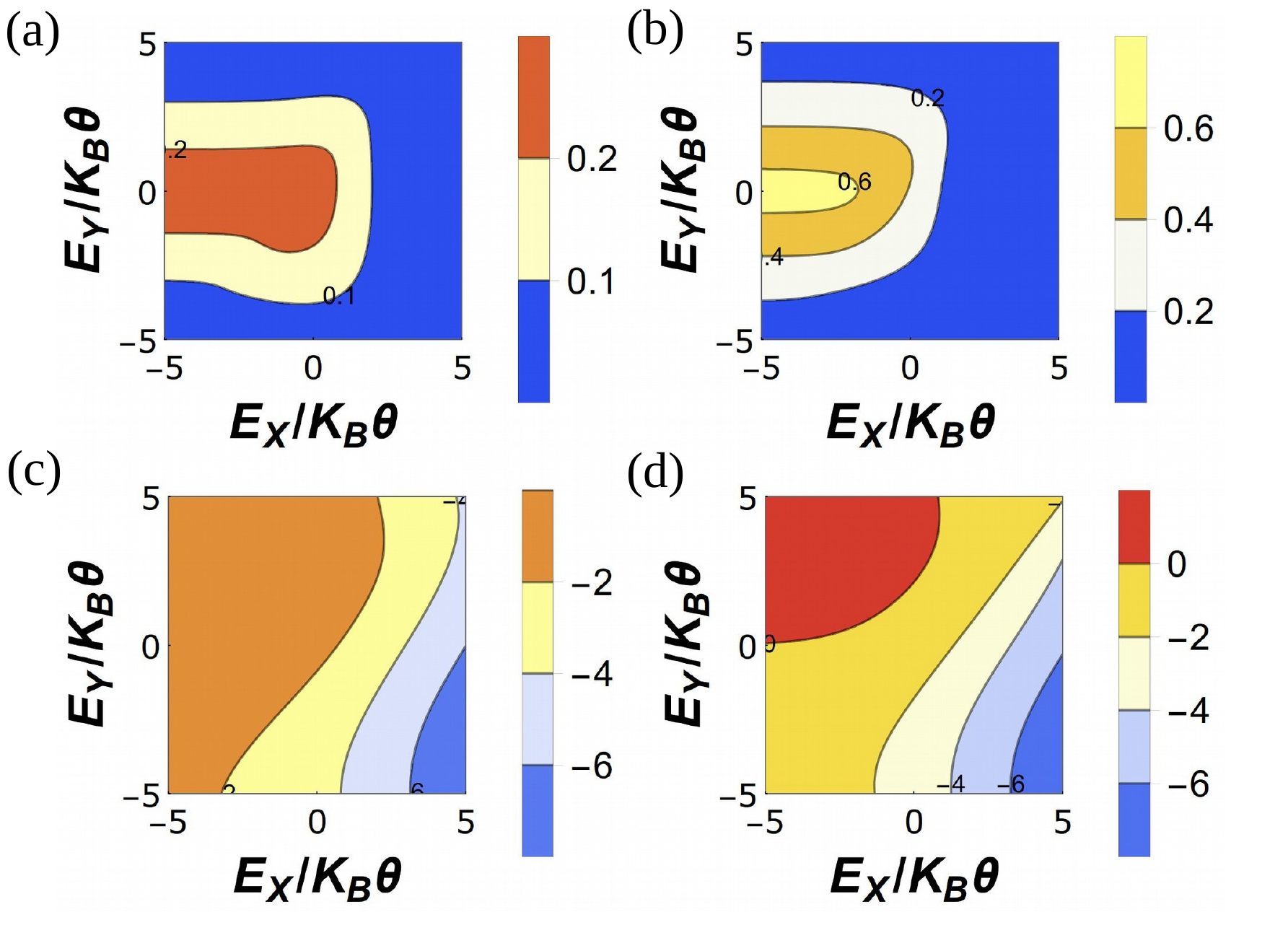}}
     \caption{(a) Spin up and (b) spin down conductances (in unit of $\frac{e^2}{h}$) are shown for QPC at constriction X and resonant tunneling at constriction Y. (c) Spin up and (d) spin down Seebeck coefficients (in unit of $\frac{k_B}{e}$) ($S^{\uparrow}$ and $S^{\downarrow}$) are shown for QPC at constriction X and resonant tunneling at constriction Y. Parameters are $\hbar \omega_0=0.1 k_B\theta$, $\Gamma=2k_B\theta$ and $\theta=0.1K$.}
     \label{con}
\end{figure}
 Since TR symmetry is preserved in a QSH system, which is also seen from the Onsager relations between the off-diagonal coefficients, we have high Peltier coefficients along with high Seebeck coefficient. A high Seebeck coefficient is a necessary condition to get a QHE with large power, a high Peltier coefficient is required condition to get a quantum refrigerator with large cooling power \cite{brandner}. 
 
\section{Results and Discussion} Our aim is to design a powerful QSH heat engine as well as a good refrigerator. For these twin purposes we need to have a large Seebeck as well as large Peltier coefficient. Seebeck and Peltier coefficients are related to the off-diagonal elements of the Onsager matrix, $L^s_{e\theta}$ and $L^s_{hV}$ respectively, as shown in Eq.~2. First we discuss the conditions required to have a powerful QSH heat engine. To have large charge power (${\mathcal{P}}^{max}_{ch}$) we need a large thermoelectric response $L^+_{e\theta}$ with small charge conductance $G_{ch}$, as in Eq.~(8). The efficiency at that charge power (see Eq.~(9)) will be large only when the thermal conductance $L^+_{h\theta}$ is small along with the condition for large power. For each of the four configurations explained before (see paragraph above Eq. (14)), we have analyzed the results. From the thermoelectric properties, maximum power and efficiency for each of these configurations we find that those properties depending on charge currents are best seen for configuration 2 (QPC at X and antidot at Y), while properties related to spin currents are best seen for configuration 1 (QPC at both X and Y). Hence, we have shown the maximum power and efficiency of charge current for configuration 2 (see Fig.~4(a,b)) and the same of spin current for configuration 1 (see Fig.~4(c,d)).
\begin{figure}
\centering {\includegraphics[width=0.4\textwidth]{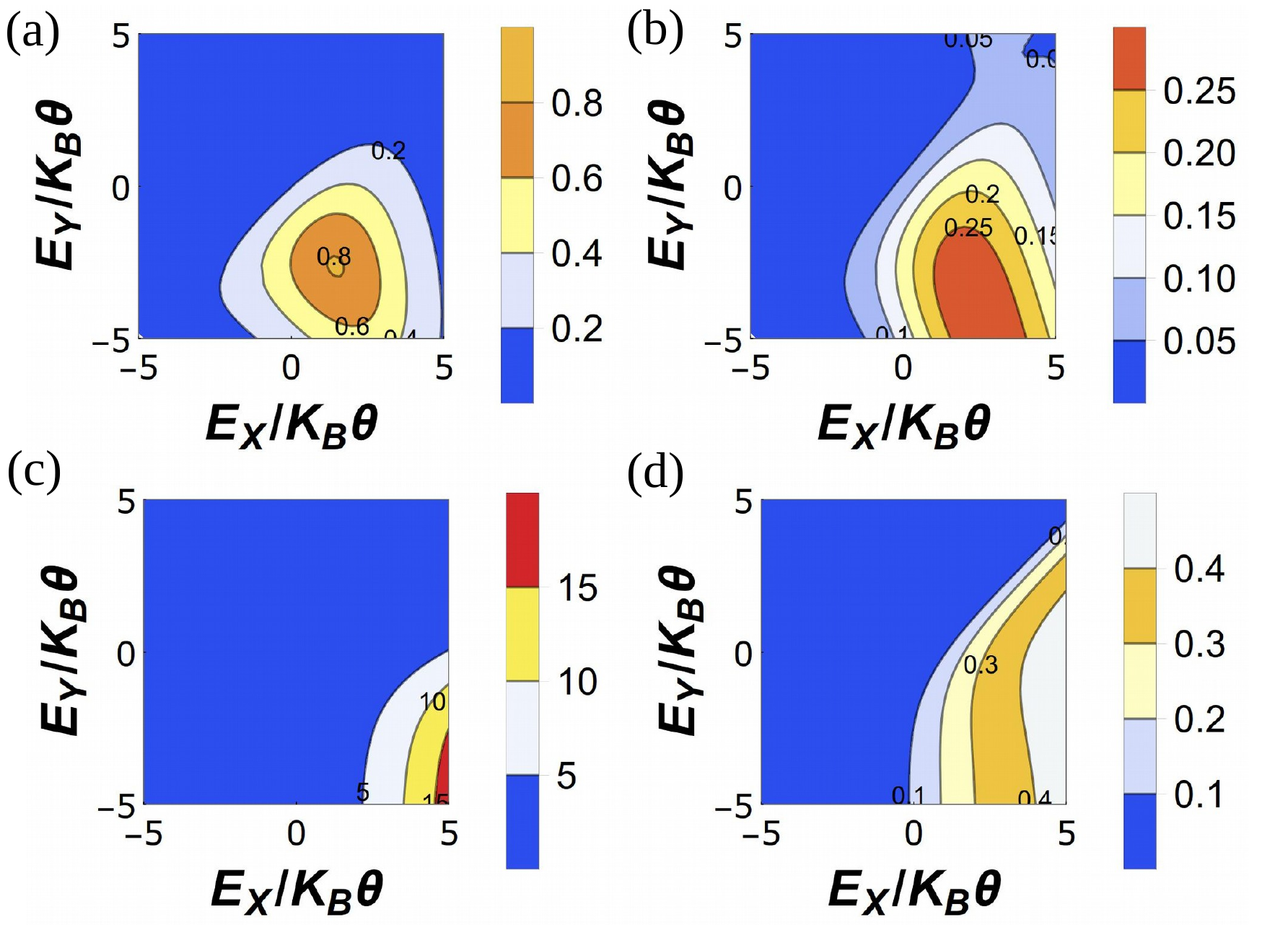}}
 \caption{ (a) Maximum power for charge currents ${\mathcal{P}}^{max}_{ch}$ in unit of $\frac{(k_B\Delta \theta)^2}{h}$ and (b) Efficiency at that power in unit of $\eta_c$ for both in configuration 2. (c) Maximum power for spin currents ${\mathcal{P}}^{max}_{sp}$ in unit of $\frac{(k_B\Delta \theta)^2}{h}$ and (d) efficiency at that power in unit of $\eta_c$ for both in configuration 1. Parameters are $\hbar \omega_0=0.1 k_B\theta$ and $\theta$=0.1K.}
 \label{pow}
\end{figure}

\subsection{ Conductance and Seebeck coefficient} For transport through QPC if, $-E_l>>\hbar\omega_{0}$ then it is open, i.e., the transmission through QPC is $1$, but if $|E_l|\leq\hbar\omega_{0}$ then it is noisy, i.e., electrons are partially transmitted through QPC, else if $E_l>\hbar\omega_{0}$ the QPC is closed. For transport through antidot, if $|E_l|>>\hbar\omega_{0}$ then it is closed, but if $|E_l|<\hbar\omega_{0}$ then it is partially open. In Fig.~\ref{con} (a, b), for configuration 2, we see that spin up and down conductances are maximum when constriction at Y is partially open, i.e., $|E_Y|\leq \hbar\omega_{o}$ and at X is open. In Fig.~\ref{con} (c,d), for the same configuration, the spin up Seebeck coefficient $|S^{\uparrow}|$ is maximum when constriction at X is closed and at Y is open. Similarly, the spin down Seebeck current $|S^{\downarrow}|$ is maximum when constriction at X is closed and Y is open. \\
\subsection{ Power and efficiency of QSH heat engine} In Fig.~\ref{pow} (a), we see the maximum charge power as large as $0.25 (k_B\Delta \theta)^2/h$  with efficiency at that power equal to  $0.8 \eta_c$ (for configuration 2), as shown in Fig.~\ref{pow} (b). We see these large power and efficiency occurs at the same parameter value where the Seebeck coefficients $|S^{\uparrow}|$ and $|S^{\downarrow}|$ are maximum, as in Eqs.~(3, 8). The power and efficiency both are maximum when constriction at X is partially open and at Y is open. The maximum power delivered by our system is double that of a quantum Hall(QH) system, due to presence of helical edge modes rather than chiral, although the efficiency generated at that maximum power is comparable to the QH system\cite{sothmann2}. The use of QSH system to design a quantum spin heat engine is only possible because of the presence of spin up/down edge modes. This is exclusive to our QSH heat engine. In Fig.~\ref{pow} (c), we see that a large spin power $15 (k_B\Delta \theta)^2/h$ is obtained in case of spin currents with efficiency at that spin power $0.4 \eta_c$ (for configuration 1), as shown in the Fig.~\ref{pow}(d). The maximum power and efficiency for spin currents are maximum when constriction at X is closed while that at  Y is open.
 \begin{figure}
\centering {\includegraphics[width=0.4\textwidth]{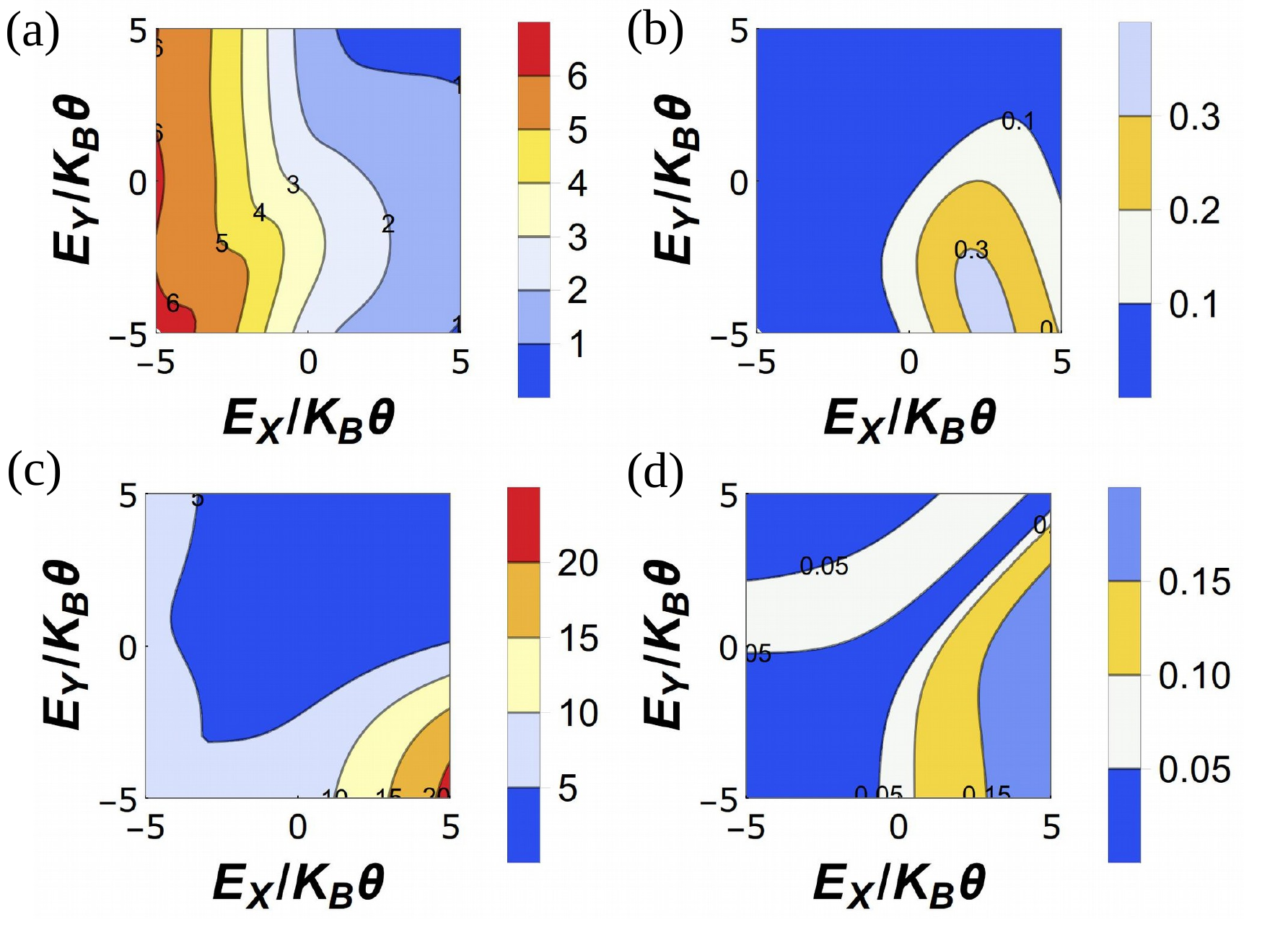}}
 \caption{(a) Maximum cooling power for charge currents $J^Q(\eta^{r,max}_{ch})$ in unit of $\frac{(k_B^{2} \theta \Delta\theta)}{h}$ and (b) maximum efficiency $\eta^{r,max}_{ch}$ in unit of $\eta^r_c$ for configuration 2. (c) Maximum cooling power for spin currents $J^Q(\eta^{r,max}_{sp})$ in unit of $\frac{(k_B^{2}\theta \Delta\theta)}{h}$ and (d) maximum efficiency $\eta^{r,max}_{sp}$ in unit of $\eta^r_c$ for configuration 1. Parameters are $\hbar \omega_0=0.1 k_B\theta$ and $\theta$=0.1K.}
 \label{ref}
\end{figure}

\subsection{Coefficient of performance(COP) and cooling power of QSH refrigerator} Next we discuss the use of the quantum spin Hall system as a charge or spin refrigerator. In Fig.~\ref{ref} (a,b), the cooling power $(J^Q(\eta^{max}_{ch}))$ (see Eq.~(11)) for charge currents of around $3.5 (k_B^{2}\theta \Delta\theta)/h$ with a COP $0.2\eta^r_c$ is observed for configuration 2. We see that the cooling power  $(J^Q(\eta^{max}_{ch}))$ is maximum when both  constrictions at X and Y are open, although the coefficient of performance for charge currents is maximum when constriction at X is partially open while at Y is open. In Fig.~\ref{ref} (c,d), the cooling power ($J^Q(\eta^{max}_{sp})$) for spin currents is shown in (c), which is around $20 (k_B^{2}\theta \Delta\theta)/h$ and maximum COP $(\eta^{max}_{sp})$of around 0.15 $\eta^r_c$ is shown in Fig. \ref{ref}(d) for configuration 2. The cooling power and COP for spin currents are maximum when constriction at X is closed and at Y is open. Again because of the preservation of TR symmetry in our system, it can act as a very good refrigerator with giant cooling power of 3.5 $(k_B^{2}\theta \Delta\theta)/h$ for charge refrigeration which is more than $150$ times than that seen in the quantum dot (QD) refrigerators (see Table II) \cite{zhang, wang}.

\section{Experimental Realization} 2D QSH samples are well known topological insulators, known for their dissipation less spin transport. These helical edge modes have been experimentally realized, see Refs. \onlinecite{josep, konig}. Though the design of a QPC in a QSH insulator is not so easy, very recently they have been experimentally realized in Ref.~\onlinecite{math}. Realization of  resonant tunneling in QSH system can be done by an antidot\cite{hub}. Thus, the experimental realization of our model would not be that difficult. { Spin power of our system can also be converted to charge power by using inverse spin Hall effect or spin valve for the system to do  electrical work as shown in Ref. \cite{Bauer}.}
{
\begin{center}
\begin{table}
\caption{How does the QSH heat engine compare with quantum Hall heat engine proposals?}
\begin{tabular}{ |p{2.1cm}|p{1.5cm}|p{1.9cm}|p{2.7cm}|}
 \hline
Heat Engines&{Maximum power ${\mathcal{P}}^{max}_{ch}$ $(k_B\Delta \theta)^2/h$}& {Efficiency at maximum power $\eta({\mathcal{P}}^{max}_{ch})$}&{Power generated in 1$cm^2$ area fabricated by nano engines}  \\ 
\hline
QH (MZI)(3T)\cite{sothmann}&0.14&0.042 $\eta_c$&0.04 Watt\\
\hline
QH (QPC) \cite{sothmann2} &0.4&0.3 $\eta_c$&0.11 Watt\\
\hline
This paper&0.8&0.28$\eta_c$&0.22 Watt\\
\hline
\end{tabular}
\end{table}
\end{center}
}
{
\begin{center}
\begin{table}
\caption{Comparison with quantum dot (QD) refrigerators}
\begin{tabular}{ |p{3.7cm}|p{2.5cm}|p{1.8cm}|}
 \hline
Quantum refrigerator& {Cooling Power $J^Q$ at maximum COP in units of $(k_B^{2}\theta \Delta\theta)/h$}&{Maximum COP in units of $\eta^r_c$}  \\ 
\hline
QD refrigerator\cite{zhang}&0.002&0.4 \\
\hline
Magnon QD refrigerator \cite{wang} &0.025&0.2 \\
\hline
This paper&3.5&0.2\\
\hline
\end{tabular}
\end{table}
\end{center}
}
\section{Conclusion} We have shown in this work that a topological insulator (QSH insulator) can work both as a charge/spin heat engine as well as a charge/spin refrigerator which uses charge/spin currents to extract heat from a cooler region of the system to dump it into a hotter region of the system. We have also compared our model with some other quantum heat engines and refrigerators in Table I and Table II respectively. In Table I, we see that the maximum output power and efficiency at that maximum charge power are much larger than the QH heat engine as in Refs.~\cite{sothmann, sothmann2}. In Table II, we see that as quantum refrigerator, the maximum charge COP of our model is comparable to other models as shown but the cooling power of our model is huge compared to other proposals. 

\section{Appendix}
Herein we provide the details of the derivation of Eqs.~(15, 16) and (17, 18) of the manuscript. Eqs.~(15, 16) relate the charge/spin current of our three terminal quantum spin Hall system to potential and thermal biases, while Eqs.~(17, 18) relate the heat current in our system to potential and thermal biases.

\subsection{{Electric charge/spin transport }}
The conduction of spin up and spin down electrons can be calculated in a three terminal quantum spin Hall (QSH) bar following Landauer-Buttiker(L-B) formalism\cite{sanchez}-
\begin{eqnarray}
I^{e,s}_i=\sum_jG^s_{ij}V_j+\sum_j L^s_{ij,e\theta}\Delta \theta_j,
\end{eqnarray} 

where, $G^s_{ij}=\frac{e^2}{h}\int_{-\infty}^{\infty} dE [M^s_i\delta_{ij}-T^s_{ij}(E)](-\frac{df}{dE})$, $L^s_{ij,e\theta}=\frac{e}{h\theta}\int_{-\infty}^{\infty} dE (E-\mu) [M^s_i\delta_{ij}-T_{ij}(E)](-\frac{df}{dE})$ with $M^s_i=$the no. of edge modes at contact `$i$' for spin `$s$' electron (in our work $M^s_i=1$ for $i=1,2,3$ and $s=\uparrow/\downarrow$), $T^s_{ij}$ is the transmission probability from terminal `$j$' to terminal `$i$' for spin `$s$' electrons, $\mu$ is the Fermi energy, `$E$' is energy of electrons and `$f$' is the Fermi-Dirac distribution, $\Delta\theta$ is the temperature bias applied at terminal $3$.  
 The spin polarized conductances $G^s_{ij}$ are related to the the constriction conductances $G_l$ with $l=X, Y$ where 
 \begin{equation}
 G_l=\frac{e^2}{h}\int_{-\infty}^{\infty} dE \mathcal{T}_l(E)(-\frac{df}{dE}),
 \end{equation}
 with $\mathcal{T}_l(E)$, the transmission probability through constriction $l=X, Y$. $T^s_{11}$ is the probability of a electron coming out of terminal 1 and going again back to the same terminal after reflection at constrictions X. Thus, $1-T^s_{11}$ implies an electron coming out of terminal $1$ but not going back into the same terminal, i.e., the transmission probability to transmit through constriction $X$ without getting scattered, which is defined by $G^s_{11}$. So, $G^\uparrow_{11}=G^\downarrow_{11}=G_X$, the constriction conductance. The conductance $G^s_{12}$ is related to the transmission probability $T^s_{12}$ of a spin `$s$' electron to transmit from terminal $2$ to $1$. $T^\uparrow_{12}$, the transmission probability of spin up electron from terminal $2$ to $1$ , shown by the blue dashed line in Fig.~1, is zero due to helical transport. We have spin up edge modes moving from left to right at the bottom edge while spin down edge modes move from right to left at top edge. Thus $G^\uparrow_{12}=0$, but $T^\downarrow_{12}$ the transmission probability for spin down electron from terminal $2$ to terminal $1$, shown by the maroon solid line in Fig.~1, is equal to the product of the transmission probabilities at constrictions X and Y because a spin down electron emitted from terminal $2$ passes the constriction Y with probability $\mathcal{T}_Y(E)$ and then constriction X with probability $\mathcal{T}_X(E)$ to enter terminal $1$. So, $T^\downarrow_{12}=\mathcal{T}_X(E)\mathcal{T}_Y(E)$ and $G^\downarrow_{12}=-J_1$ (minus sign is due to the current flowing in a clockwise direction), where 
 \begin{equation}
 J_n=A_n\int_{-\infty}^{\infty}  dE (E-\mu)^{n-1} \mathcal{T}_X(E)\mathcal{T}_Y(E)(-\frac{df}{dE}),
 \end{equation}
with $A_1=\frac{e^2}{h}$, $A_2=\frac{e}{\theta h}$, $A_3=\frac{1}{\theta h}$. The thermopower $S_l$ generated across the QPC's/antidots at constriction `$l$' is defined as-
\begin{equation}
 S_l=\frac{e}{\theta hG_l}\int_{-\infty}^{\infty}  dE (E-\mu)(-\frac{df}{dE})\mathcal{T}_l(E)
 \end{equation}
Similarly, $L^\uparrow_{13,e\theta}$ depends on the transmission probability $T^\uparrow_{13}$ of a spin up electron from terminal $3$ to terminal $1$ (see the expression for $L^s_{ij,e\theta}$ below Eq.~(19) ). The spin up electron emitted from terminal $3$ enters terminal $1$ after passing through the constriction X, thus, $T^\uparrow_{13}=\mathcal{T}_X(E)$ and $L^\uparrow_{13,e\theta}=-G_XS_X$. The rest of the conductances $G_{i,j}^{s}$ and thermoelectric responses, $L^s_{ij,k}$'s too depend on the transmission probability from terminal $j$ to $i$ in a similar fashion. Thus, electric current and voltages at the three terminals are related as follows-
\begin{widetext}
\begin{eqnarray}
  \left( {\begin{array}{c}
   I^{e,\uparrow}_1  \\
   I^{e,\downarrow}_1  \\
    I^{e,\uparrow}_3\\
    I^{e,\downarrow}_3\\
  \end{array} } \right)= \left( {\begin{array}{cccc}
 G_X&0 & -G_X &-G_XS_X \\
   G_X& -J_1&-G_X+J_1&-G_XS_X+J_2 \\
   -G_X+J_1&-G_Y&G_X+G_Y-J_1&G_XS_X+G_YS_Y-J_2\\
   -G_X&-G_Y+J_1&G_Y+G_Y-J_1&G_XS_X+G_YS_Y-J_2\\
  \end{array} } \right) \times\left( {\begin{array}{c}
   V_1  \\
  V_2  \\
  V_3\\
  \Delta \theta\\
  \end{array} } \right)
\end{eqnarray}   
\end{widetext}
Since the third terminal is an ideal voltmeter, electric charge current through this terminal is zero $I^e_{ch,3}=0$ and as terminal $2$ is grounded, $V_2=0$. So, the total electric current $I^e_{ch,3}=I^{e,\uparrow}_3+I^{e,\downarrow}_3=0\Rightarrow V_3=\frac{2G_X-J_1}{2(G_X+G_Y-J_1)}V_1-\frac{G_XS_X+G_YS_Y-J_2}{G_X+G_Y-J_1}\Delta \theta$. Substituting this value in Eq.~(23), we get-
\begin{eqnarray}
 I^{e,\uparrow}_1=G^\uparrow V_1+L^{\uparrow}_{e\theta}\Delta \theta,\quad\text{and}\quad
  I^{e,\downarrow}_1=G^\downarrow V_1+L^{\downarrow}_{e\theta}\Delta \theta,
\end{eqnarray}
wherein,
\begin{eqnarray}
G^\uparrow&=&\frac{G_X(2G_Y-J_1)}{2(G_X+G_Y-J_1)}, \qquad G^\downarrow=\frac{2G_XG_Y+G_XJ_1-J_1^2}{2(G_X+G_Y-J_1)},\nonumber\\
L^{\uparrow}_{e\theta}&=&\frac{G_XG_Y}{G_X+G_Y-J_1}(S_Y-S_X)+\frac{G_X}{G_X+G_Y-J_1}(J_1S_X-J_2),\nonumber\\
L^{\downarrow}_{e\theta}&=&\frac{G_XG_Y}{G_X+G_Y-J_1}(S_Y-S_X)-\frac{G_Y}{G_X+G_Y-J_1}(J_1S_Y-J_2),\nonumber\\
\end{eqnarray}
for spin up and down electric currents. Eqs.~(24,25) are Eqs.~(15,16) of manuscript.

{\subsection{Heat transport} }
For a multi-terminal setup with thermoelectric transport, the heat currents using Landauer-Buttiker formalism are given as follows\cite{sanchez}-
\begin{eqnarray}
I^{h,s}_i=\sum_jL^s_{ij,hV}V_j+\sum_j L^s_{ij,h\theta}\Delta \theta_j,
\end{eqnarray} 
where $L^s_{ij,hV}=\frac{e}{ h}\int_{-\infty}^{\infty} dE (E-\mu)(-\frac{df}{dE})[M^s_i\delta_{ij}-T_{ij}(E)]$ and $L^s_{ij,h\theta}=\frac{1}{\theta h}\int_{-\infty}^{\infty} dE (E-\mu)^2[M^s_i\delta_{ij}-T_{ij}(E)](-\frac{df}{dE})$. The Peltier term $L^s_{11,hV}$ depends on the probability $(1-T^s_{11})$ (see the expression for $L^s_{ij, hV}$ ) for spin $s$ electrons. $T^s_{11}$ is the probability of a spin `$s$' electron emitted from terminal 1 to again go back to same terminal, after getting reflected at the constrictions X. For spin up electron, probability $(1-T^\uparrow_{11})$ defines the transmission for spin up electron coming out of terminal 1 and not going back to the same terminal (see the blue dashed line in Fig.~1), i.e., after coming out of terminal 1, it is transmitted through the constriction X, so $(1-T^\uparrow_{11})=\mathcal{T_X(E)}$. Thus, $L^\uparrow_{11,hV}=L^\downarrow_{11,hV}=G_XS_X\theta=\frac{e}{ h}\int dE (E-\mu)(-\frac{df}{dE})\mathcal{T}_X(E)$. Similarly, the thermal conductance $L^\uparrow_{13,h\theta}$ depends on the transmission of thermal current from terminal 3 to 1, i.e. on the transmission function $T^\uparrow_{13}=\mathcal{T}_X(E)$ (see the blue dashed line in Fig.~1) for spin up electron, so $L^\uparrow_{13,h\theta}=-N_X$, where
\begin{eqnarray}
N_l=\frac{1}{\theta h}\int_{-\infty}^{\infty}  dE \mathcal{T}_l(E)(E-\mu)^2(-\frac{df}{dE}),
\end{eqnarray}
 is the thermal conductance across the QPC/antidot at constriction `$l$' with $l=X,Y$. Each of the entries in matrix (Eq.~(28)) can be explained in this way.
\begin{widetext}
\begin{eqnarray}
  \left( {\begin{array}{c}
    I^{h,\uparrow}_1\\
    I^{h,\downarrow}_1\\
      I^{h,\uparrow}_2\\
    I^{h,\downarrow}_2\\
      I^{h,\uparrow}_3\\
    I^{h,\downarrow}_3\\
  \end{array} } \right)= \left( {\begin{array}{cccc}
 G_XS_X\theta&0 & -G_XS_X\theta &-N_X \\
   G_XS_X\theta& -J_2\theta&-(G_XS_X-J_2)\theta&-(N_X-J_3) \\
   -J_2\theta&G_YS_Y\theta&-(G_YS_Y-J_2)\theta&-(N_Y-J_3)\\
   0&G_YS_Y\theta&-(G_YS_Y)\theta&-N_Y\\
   -(G_XS_X-J_2)\theta&-G_YS_Y\theta&(G_XS_X+G_YS_Y-J_2)\theta&(N_X+N_Y-J_3)\\
    -G_XS_X \theta&-(G_YS_Y-J_2)\theta&(G_XS_X+G_YS_Y-J_2)\theta&(N_X+N_Y-J_3)\\
  \end{array} } \right) \times\left( {\begin{array}{c}
   V_1  \\
  V_2  \\
  V_3\\
  \Delta \theta\\
  \end{array} } \right)
\end{eqnarray}   
\end{widetext}

In our setup, we need only the heat current $I^{h,\uparrow}_3$ and $I^{h,\downarrow}_3$ at terminal $3$ in terms of the potential bias and thermal bias, by putting the value of $V_3$ (as derived in Appendix A) in terms of $V_1$ and $\Delta \theta$ we get-
\begin{eqnarray}
I^{h,\uparrow}_3=L^\uparrow_{hV}V_1+L^\uparrow_{h\theta}\Delta \theta,\nonumber\\
I^{h,\downarrow}_3=L^\downarrow_{hV}V_1+L^\downarrow_{h\theta}\Delta \theta,
\end{eqnarray}
where, 
\begin{widetext}
\begin{eqnarray}
L^\uparrow_{hV}&=&\theta L^\downarrow_{e\theta}=\frac{\theta G_XG_Y}{G_X+G_Y-J_1}(S_Y-S_X)-\frac{\theta G_Y}{G_X+G_Y-J_1}(J_1S_Y-J_2), \nonumber\\
L^\downarrow_{hV}&=&\theta L^\uparrow_{e\theta}=\frac{\theta G_XG_Y}{G_X+G_Y-J_1}(S_Y-S_X)+\frac{\theta G_X}{G_X+G_Y-J_1}(J_1S_X-J_2), \nonumber\\
L^{\uparrow}_{h\theta}&=&L^{\downarrow}_{h\theta}=(N_1+N_2-J_3)-\theta\frac{(G_XS_X+G_YS_Y-J_2)^2}{(G_X+G_Y-J_1)}.
\end{eqnarray}
\end{widetext}
From Eq.~(30) we see that $(L^\uparrow_{hV}+L^\downarrow_{hV})=\theta (L^{\uparrow}_{e\theta}+L^{\downarrow}_{e\theta})$, which implies that the TR symmetry is preserved in three terminal QSH systems unlike in three terminal QH systems, see Ref.~\cite{sothmann2}. Eqs.~(29, 30) are Eqs.~(17, 18) of the main manuscript.

\end{document}